\documentclass{article}
\usepackage[utf8]{inputenc}
\usepackage{authblk}
\usepackage{hyperref}
\usepackage{bm}
\usepackage{setspace}
\usepackage{braket}
\usepackage{amsmath}
\usepackage{amssymb}
\usepackage{physics}
\usepackage{mathtools}
\usepackage{xcolor}
\usepackage{graphicx}

\title {Revisiting Nancy Cartwright's Notion of Reliability: Addressing Quantum Devices' Noise}
\author{Galina Weinstein}
\affil{\normalsize Reichman University, The Efi Arazi School of Computer Science, Herzliya; University of Haifa, The Department of Philosophy, Haifa, Israel.}

\begin{document}

\maketitle

\begin{abstract} 
This paper serves as an addendum to my previously published work, which delves into the experimentation with the Google Sycamore quantum processor under the title "Debating the Reliability and Robustness of the Learned Hamiltonian in the Traversable Wormhole Experiment." In the preceding publication, I extensively discussed the quantum system functioning as a dual to a traversable wormhole and the ongoing efforts to discover a sparse model that accurately depicts the dynamics of this intriguing phenomenon.
In this paper, I bring to light an important insight regarding applying Nancy Cartwright's ideas about reliability and reproducibility, which are deeply rooted in classical scientific practices and experiments. I show that when applied to the realm of quantum devices, such as Google's Sycamore quantum processor and other Noisy Intermediate-Scale Quantum (NISQ) devices, these well-established notions demand careful adaptation and consideration.
These systems' inherent noise and quantum nature introduce complexities that necessitate rethinking traditional perspectives on reliability and reproducibility. In light of these complexities, I propose the term "noisy reliability" as a means to effectively capture the nuanced nature of assessing the reliability of quantum devices, particularly in the presence of inherent quantum noise.
This addendum seeks to enrich the discussion by highlighting the challenges and implications of assessing quantum device reliability, thereby contributing to a deeper understanding of quantum experimentation and its potential applications in various domains.
\end{abstract}

\newpage
\section{Introduction}

This paper serves as an addendum to my previously published work \cite{Weinstein}, wherein I conducted an in-depth exploration of the experimentation involving the Google Sycamore quantum processor. In that earlier publication, I extensively discussed the quantum system that functions as a dual to a traversable wormhole and the dedicated endeavors to identify a sparse model that accurately captures the dynamics inherent in such a phenomenon.

The preceding paper intricately explored the properties of the quantum system, specifically focusing on the sparse model known as the learned Hamiltonian, carefully crafted to simulate the dynamics of traversable wormholes faithfully. The discussion in the previous publication encompassed a comprehensive analysis of the sparse model, addressing its criticisms and shedding light on the challenges arising from the background noise within the Sycamore device. Furthermore, the perspective of reliability and robustness was meticulously applied, critically examining the experiment's fidelity in confronting these challenges.

However, my journey remains incomplete without delving deeper into the persistent quandary posed by noise in quantum devices. In this present endeavor, I aim to explore the depths of this predicament, viewing it through Nancy Cartwright's philosophical lens of reliability.

Subsequently, I explore the intricate relationship between noise, reliability, and the philosophical underpinnings that intertwine them. This paper is a companion to my previous work, offering a more profound reflection on the challenges and implications of noise and inviting contemplation on the fundamental nature of reliability in the quantum domain.

In the realm of quantum systems, where inherent noise plays a crucial role in the experimental setting, the conventional boundaries of reliability and reproducibility become indistinct. Within this context, I navigate the nuances of assessing the reliability of quantum devices, searching for a term that encapsulates the essence of this dynamic interplay. The proposed term - "noisy reliability" - emerges, capturing the essence of my quest for a comprehensive validation of experimentation within the realm of quantum noise. It effectively reflects the acknowledgment of uncertainty in quantum mechanics while seeking a new understanding of quantum device performance.

In the forthcoming sections, I delve into Cartwright's thesis on the tangle of science and reliability in Section \ref{1}, then explore the intrinsic relationship between reliability and reproducibility in classical physics within Section \ref{2}. As I venture further, Section \ref{3} unravels the notion of reliability in the context of quantum systems and devices, highlighting quantum noise's challenges. Ultimately, my journey culminates in section \ref{4}, in coining the term "noisy reliability" – a term that encapsulates the call for a new perspective on validation in the realm of inherently noisy instruments.

\newpage
\section{Cartwright on Reliability} \label{1}

Cartwright argues that many scientific products we need to evaluate are not focused on "truth," and therefore, we should shift our perspective to assess their "\emph{in situ} reliability." Reliability refers to the ability of a scientific theory or model to produce accurate predictions in a specific context. 
Cartwright advocates for a shift in focus towards the reliability of science's panoply of outputs science produces,  \emph{the tangle of science}. This shift entails moving from a general perspective to a more specific and detailed examination.

Cartwright's notion of the \emph{tangle of science} is an extensive and diverse network comprising various components that can be utilized in diverse ways to serve different purposes. It encompasses not only theoretical and experimental elements but also models, concepts, measurement definitions, methods, bridging principles, procedures, instruments, practices, concept development and validation, case studies, coding, methods of approximation, methods of inference, measures, evaluations, devices, statistical techniques, data collection, analysis, curation, production, preservation, classification, dissemination, non-experimental studies, narratives, plans, science-informed policies, and numerous other products of scientific inquiry. 
Each of these products carries equal importance in the scientific landscape.
Particularly significant is how these products interconnect and interweave, with their collective synergy contributing to the success of any scientific endeavor \cite{Cartwright1}, \cite{Cartwright3}. 

In the context of the experiment discussed in my previous paper \cite{Weinstein}, the approach advocated by Cartwright necessitates the evaluation of the reliability of specific components within scientific practices. These components include the quantum circuit and learned Hamiltonian, the sparsification process (see \cite{Jafferis}), and the reliability of the Sycamore quantum processor.

The quantum circuit is composed of the following elements: researchers prepared the thermofield double (TFD) state $\ket{\text{TFD}}$, which consists of the eigenstates of the left and right quantum systems. They then created a maximally entangled state between registers $P$ and $Q$. This step ensures high entanglement between these two registers. Afterward, at the time $t = t_0$, they applied a SWAP operation between registers $Q$ and $L$ to insert the qubit into the wormhole, creating an entangled connection between them. At the time $t = 0$, they applied an interaction between the left and right quantum systems, which is carried out between register $R$ and register $L$. This interaction allows the qubit to interact with the left and right quantum systems. Finally, at the time $t = t_1$, they applied another SWAP operation between registers $R$ and $T$ to extract the qubit from the wormhole, transferring the qubit's entanglement from register $L$ to register $T$. The mutual information $I_{PT}$ is measured between the registers $P$ and $T$.\footnote{In the above quantum circuit protocol, $I_{PT}$ measures the entanglement between register $P$, where a maximally entangled state was prepared, and register $T$, where the qubit was extracted from the wormhole. When we measure register $T$, the measurement outcome collapses the quantum state of register $T$ to one of the basis states. This measurement outcome will be correlated with the quantum state of register $P$ due to the prior entanglement achieved between registers $P$ and $T$ through the interaction and SWAP operations. After executing the protocol, the measured mutual information $I_{PT}$ will quantitatively measure the entanglement between these two registers \cite{Jafferis}, \cite{Zlokapa}.}

Cartwright emphasizes that by focusing on reliability rather than exclusively on truth, we can gain deeper insights into the effectiveness of scientific practices and their outputs. She suggests that the concern for reliability is essential because truth, confirmation, and warrant are contingent upon it. While she advocates for investigating the attributes that make these scientific components reliable and trustworthy, she does not dismiss the importance of truth altogether. However, she points out that assuming the truth of a scientific claim without good reason and support for its reliability puts us on shaky ground.

The shift in focus proposed by Cartwright encourages us to move away from evaluating the truth of general principles and instead concentrate on the reliability of various scientific products. Many of these products do not apply to being labeled as "true" or "false." Evaluating their reliability becomes crucial, raising the question of what they are reliable for, which is often overlooked. Overall, Cartwright's advice calls for a comprehensive assessment of the reliability of scientific outputs rather than solely seeking truth. By focusing on reliability rather than exclusively on truth, researchers can gain a deeper understanding of the effectiveness and trustworthiness of the experiment's results \cite{Cartwright1}, \cite{Cartwright2}, \cite{Cartwright3}, \cite{Cartwright4}, \cite{Galison}.

\section{Reliability and Reproducibility} \label{2}

In her paper titled "Replicability, Reproducibility, and Robustness," Cartwright asserts her lack of confidence in the reliability of instruments when issues arise due to the implementation of a model rather than inherent problems with the instruments themselves. Even if the instruments are highly accurate in isolation, the process of implementing the model can magnify small errors and uncertainties, resulting in less reliable overall predictions \cite{Cartwright5}.

According to Cartwright, the reliability of scientific practices and their outputs is considered essential because truth and confirmation rely on this aspect.
Suppose there is an issue with the reliability of the sparse model, referred to as the learned Hamiltonian, which is designed to represent traversable wormhole dynamics. In such a scenario, the implications extend to the accuracy of the underlying model of a holographic semi-classical wormhole. The learned Hamiltonian plays a pivotal role in the experiment, representing the quantum system and governing its behavior. Hence, if the Hamiltonian is unreliable or inaccurate, it leads to erroneous predictions and unreliable outcomes within the experiment. I extensively addressed this in my prior paper \cite{Weinstein}.
Therefore, the reliability of the learned Hamiltonian becomes critical, as it significantly impacts the overall reliability of the experimental results. The trustworthiness of the Hamiltonian has direct implications for the validity of the underlying model and the conclusions drawn from the experiment. Any uncertainties or errors in the Hamiltonian have the potential to propagate throughout the experiment, casting doubt on the veracity of the claims and principles based on its outcomes.

Cartwright emphasizes the importance of reliability for obtaining accurate and trustworthy results, especially in cases where experiments cannot be easily reproduced. She illustrates her point by referring to the Gravity Probe B experiment, a substantial endeavor headed by Francis Everritt and a sizable team of researchers. 

Gravity Probe B, launched in 2004, is a space experiment utilizing cryogenic gyroscopes in Earth's orbit. In the Gravity Probe B project, the primary objective was to observe and quantify a relativistic frame-dragging precession phenomenon, which is a subtle effect predicted by Einstein's general theory of relativity. To achieve this high-precision measurement, the project team constructed specialized gyroscopes that exhibit uniform and synchronous rotation. 
Even slight variations in the manufacturing process of the gyroscopes could adversely affect the precision of the measurements they aimed to conduct. Therefore, the team opted for fused quartz as the gyroscope material due to its ability to be uniformly produced. This ensured that all components of the gyroscopes were identical, without any discrepancies.
Maintaining this uniformity is paramount because any variations in density within the gyroscopes could lead to additional cyclic wobbling, which is beyond control and prediction. This wobbling could subsequently hinder the precise measurement of the relativistic effects, potentially obscuring their accurate observation.

The project's design and implementation took $45$ years due to the complexity and importance of the observation. The reason for such an extended period is to ensure that the design of the experiment is highly reliable and trustworthy.
However, uniformity alone did not suffice; just making the gyroscopes uniform is insufficient. The experiment's success hinged on the team carefully considering potential external influences and disturbances that could confound the measurements. Factors such as external forces and interference were meticulously considered during the experimental design phase. This encompassed strategies to minimize these factors' impact and enhance the visibility of the relativistic precession within the experimental results.

It is noteworthy that the overarching goal of the Gravity Probe B project was of significant magnitude. It sought to substantiate a fundamental aspect of Einstein's theory of relativity by investigating the interaction between a rotating gyroscope and the curved fabric of space-time. The ultimate aim was to experimentally validate Einstein's theory's predictions by examining spinning gyroscopes \cite{Cartwright7}.
However, as the team pursued their primary objective, they stumbled upon an additional discovery that proved more practical and less abstract. This newfound insight revolves around a general idea that can be effortlessly articulated in a typical, straightforward manner. The concept proposes that any gyroscope made of fused quartz, supported by magnets, coated with a super-thin layer of superfluid, read using a special detector, kept at cryogenic temperatures, and set into rotation in the depths of space, will exhibit predictable behavior. Specifically, these gyroscopes will precess at a specific rate that can be calculated and forecasted.
Thus, while the primary aim of the Gravity Probe B project is to test Einstein's theory, it also yielded valuable information about how gyroscopes with the mentioned specific characteristics behave under particular conditions. This practical discovery can be expressed as a general rule applicable to gyroscopes possessing these unique attributes \cite{Cartwright7}.

Cartwright expresses confidence in the instrument's reliability, emphasizing that being extremely certain about its accuracy is crucial since it will likely provide a unique opportunity to gather data \cite{Cartwright5}. In 1989 Cartwright wrote, "The GP-B experiment is a good example. It may take twenty years, but in the end, it should provide an entirely reliable test for the effects of space-time curvature, a test as stringent as any empiricist could demand" \cite{Cartwright6}.

\section{Reliability of Quantum Devices} \label{3}

Now, I will apply Cartwright's ideas to the field of noisy intermediate-scale quantum (NISQ) computers such as Google's Sycamore. The learned sparse SYK model that retains critical attributes of the traversable wormhole (see \cite{Jafferis}) was executed utilizing the Sycamore processor. A blog post on Google's website stated, "The Google Sycamore processor is among the first to have the fidelity needed to carry out this experiment" \cite{Zlokapa1}. In other words, it is argued that this high fidelity indicates that the Sycamore processor can maintain the coherence and stability of quantum states throughout the computation, reducing the impact of noise and errors. As a result, the Sycamore processor can perform the wormhole experiment with great reliability and accuracy.

The inherent noise in quantum computations is a significant challenge in the field of quantum computing and impacts the reliability of the Sycamore. Quantum gates are susceptible to errors stemming from \emph{decoherence} (interactions with the external environment that introduce errors in quantum gate operations), \emph{gate crosstalk} (where the operation of one gate affects the intended operation of another nearby gate), and \emph{imprecise calibration} of the quantum gates (that can result in inaccuracies during gate operations).    
Additionally, quantum measurements can introduce noise and disturb the quantum state (\emph{readout errors}), affecting subsequent computations. Noise during the measurement can introduce errors in the final results. Generally, quantum systems inherently exhibit quantum noise due to the probabilistic nature of quantum mechanics. This inherent noise can affect the accuracy of gate operations. 
Sycamore's noisy nature makes it difficult to express extreme confidence in its reliability and accuracy, as small errors and uncertainties can accumulate during computations. This can result in less reliable overall outcomes, especially for complex calculations.
In addition to the model implementation potentially amplifying minor inaccuracies and leading to less dependable predictions, it is crucial to acknowledge that the device is susceptible to noise and errors. Noise and decoherence introduce uncertainties that limit the reliability of the Sycamore processor. 

As quantum technology advances and error-correction techniques improve, it is expected that the reliability and accuracy of quantum computations will improve over time.
But it is important to acknowledge that evaluating the reliability of the Sycamore quantum processor poses additional challenges due to quantum sensitivity and noise, which raises doubts about the trustworthiness of a single simulation run. Due to the presence of noise, we cannot have complete confidence in the reliability of the processor. 

In the case of Gravity Probe B and similar experiments involving classical physical systems, noise can emerge from various sources, including environmental factors, instrument imperfections, and external disturbances. These noise sources can introduce uncertainties and deviations in the measurements and outcomes of such experiments. These noise sources can affect measurements’ reliability and reproducibility, but they are often more predictable and manageable than the inherent noise present in quantum systems.
In Google's Sycamore, noise is inherent. Quantum states are highly delicate and can be easily disrupted by their surroundings, leading to errors in quantum computations. This intrinsic noise poses significant challenges to achieving reproducibility and reliability, as even small disturbances can lead to unpredictable and uncontrollable effects on the quantum states being manipulated.

Cartwright's theory emphasizes the importance of reliability for confirming scientific truths and obtaining accurate results. In the case of the Sycamore, the challenges posed by quantum noise can hinder our ability to achieve the same level of confidence and reliability as in macroscopic experiments like Gravity Probe B. The inherent noise in NISQ devices introduces a level of unpredictability that can make it difficult to obtain consistent outcomes across multiple runs of an experiment, affecting the results' reliability.
In other words, the inherent quantum noise is a fundamental challenge distinguishing quantum systems from traditional macroscopic experiments like Gravity Probe B.

\section{Noisy reliability} \label{4}

The inherent noise in the Google Sycamore can hinder the ability to reproduce experiments, which is a crucial aspect of scientific validation. Reproducibility allows other researchers to verify experimental results and build upon them independently. However, the inherent noise in the Sycamore makes it challenging to reproduce the same computation and obtain consistent outcomes. Reliability is intimately tied to reproducibility in scientific experiments. The presence of noise and its challenges to reproducibility in the context of the Sycamore can complicate assessing its reliability.

Suppose someone intends to execute the sparsified SYK model (which, as asserted by its creators, maintains fundamental attributes of traversable wormhole physics) on an alternate quantum device; the inherent noise within the NISQ device might result in disparities during the experiment's simulation, potentially leading to distinct experimental outcomes. Reproducing experiments across different quantum processors is not straightforward due to inherent noise; different quantum processors have varying noise levels due to differences in their qubit technologies and environmental conditions. Even if two quantum processors have similar architectures and qubit counts, they may still exhibit differences in performance and ability to reproduce the same model precisely.

I argue that Cartwright's notion of reliability needs to be adapted when applied to quantum instruments, particularly quantum processors. 
The inherent quantum nature and noise in quantum systems introduce complexities that may require rethinking traditional notions of reliability. Assessing the reliability of quantum devices, such as quantum processors, involves considering the impact of quantum noise and understanding the interplay between quantum states, measurements, and computations. This calls for a more nuanced perspective on reliability in the context of quantum computing.

In quantum computing, noisy simulating is the process of simulating a quantum system in the presence of noise. These simulations involve incorporating noise models into the algorithms and circuits being simulated. In a similar vein, I propose the term "noisy reliability."
I suggest that the term "noisy reliability" effectively captures the nuanced nature of assessing the reliability of quantum devices, especially in inevitable quantum noise. This term reflects the understanding that while perfect reproducibility may be challenging to achieve in the noisy quantum realm, researchers can still work to establish a realistic assessment of the reliability and validity of the outcomes.  
"Noisy reliability" acknowledges that noise is an inherent part of quantum systems and that researchers must navigate this uncertainty while striving to extract meaningful and accurate information from quantum experiments. It signifies a pragmatic approach that considers the limitations imposed by quantum noise while still aiming to obtain reliable and valuable results, even though quantum systems may not always produce the same results upon repeated measurements due to inherent quantum noise.

In the field of quantum computing, where noise is a fundamental challenge, the concept of "noisy reliability" captures the essence of how researchers work to understand, characterize, and mitigate noise effects to achieve trustworthy and informative outcomes. This term underscores the importance of acknowledging and addressing noise while still striving to establish a reliability level appropriate within the context of quantum systems.

\section*{Acknowledgement}

\noindent This work is supported by ERC advanced grant number 834735.

\end{document}